\documentclass[12pt,a4paper]{article}

\usepackage{amssymb}
\usepackage{setspace}
\usepackage{indentfirst}
\begin{document}

\begin{titlepage}
\begin{flushright}
KEK-TH-2091
\end{flushright}

\begin{center}

\vspace*{20mm}

{\Large\bf
\begin{spacing}{1.2}
A Possibility of Lorentz Violation in the Higgs Sector
\end{spacing}
}

\vspace*{30mm}

{\large
Satoshi Iso${}^{\; a,b}$ and Noriaki Kitazawa${}^{\; c}$
}
\vspace{10mm}

{${}^a$\sl\small KEK Theory Center, High Energy Accelerator Research Organization (KEK),\\ }
{${}^b$\sl\small Graduate University for Advanced Studies (SOKENDAI),\\
 Tsukuba, Ibaraki 305-0801, Japan \\}
\vspace{8pt}

{${}^c$\sl\small Department of Physics, Tokyo Metropolitan University,\\
 Hachioji, Tokyo 192-0397, Japan \\ }
\vspace{8pt}

\vspace*{20mm}

\begin{abstract}
In string theory, a scalar field often appears as a moduli  
of a geometrical configuration of D-branes
in higher dimensional space. 
In the low energy effective theory on D-branes, 
the distance between D-branes is translated into 
 the energy scale of the gauge symmetry breaking. 
In this paper, we study a phenomenological consequence of 
 a possibility that the Higgs field is such a moduli field 
 and the D-brane configuration is stabilized by a stationary 
motion, in particular, revolution of D-branes on which we live. 
Then, due to the Coriolis force, Higgs mode is mixed with the angular fluctuation of branes and the
 Lorentz symmetry is violated in the dispersion relation of the Higgs boson. 
The Higgs boson mass measurements at LHC experiments give an upper bound  $\sim {\cal O}(0.1)$ GeV
for the angular frequency of the revolution of D-branes. 
\end{abstract}

\end{center}
\end{titlepage}

\section{Introduction}
\label{sec:introduction}

Various configurations of D-branes 
 have been investigated to construct realistic models of particle physics 
 (for review see \cite{Blumenhagen:2006ci,Ibanez:2012zz} and references therein), 
and an important issue is how to stabilize such configurations: moduli stabilization. 
In most cases, static configurations are considered because they
preserve some supersymmetries and thus the configurations are stable.
In other situations, D-brane configurations in motion 
have been also discussed, e.g. in the context of the early cosmology,  
especially as a stringy realization of inflationary universe
 \cite{Dvali:1998pa,Kehagias:1999vr,Silverstein:2003hf,Easson:2007dh}.
In a previous paper  \cite{Iso:2015mva} we  pointed out a possibility
to utilize a stationary configuration, in particular a revolution of D-branes, 
as a possible mechanism for the electroweak symmetry breaking.
The distance between D-branes can be stabilized at least classically by the revolution.
Though explicit physical models have not been proposed yet, it is clear, however,
 we need to evade two general problems:
 destabilization by emissions of massless particles from D-branes and 
 violation of the Lorentz invariance on D-branes. 
 
In this paper we investigate Lorentz violation in the dispersion relation 
of the Higgs field 
{\it under an  assumption} that Higgs field is realized as a moduli 
field corresponding to the distance between D-branes 
that are revolving around each other with an angular frequency $\omega_0$. 
We concentrate on the Lorentz violation itself 
and leave explicit model constructions or the 
stabilization mechanism of the revolving motion for future investigations\footnote{
Such a Higgs field typically belongs to adjoint representation in gauge group,
 but some mechanisms to obtain Higgs field in fundamental representation 
 have been investigated, e.g. in \cite{Antoniadis:2000tq,Kitazawa:2012hr},
 which may be applied to our case.
A possibility of D0-brane bound states has been
 discussed in \cite{Kabat:1996cu, Danielsson:1996uw}, and
calculations of open-string one-loop amplitudes between revolving D-branes
are given in \cite{Iso-Ohta-Suyama}. 
But many issues, such as stabilization of stationary configurations 
or associated supersymmetry breaking, are not yet understood and wait for further investigations.}.
The following investigation is generic in the sense that 
 it is independent of further details of underlying model constructions or the dynamics of stabilization.
The only necessary assumption is 
 that the Higgs field is represented as the open string stretching between revolving D-branes.

The Lorentz violation in particle physics are both intensely and extensively studied 
\cite{LV} but most of the studies are performed in the QED, gravity, and some of the 
standard model (SM) particles,  but not much in  the Higgs sector (see \cite{LV-Higgs}). 
In string theory,  the Higgs field may be qualitatively different from other fields since
the Higgs field, the only scalar field in the standard model, may have a geometrical origin. 
Thus we cannot exclude a possibility
that the Lorentz violation occurs only in the Higgs sector. 
In a scenario of the moduli stabilization based on a stationary motion of D-brane configurations,
 we find that the violation of Lorentz symmetry appear 
in the dispersion relations of a D-brane moduli field, 
 which is assumed to be identified as the Higgs field.  
Lorentz violation does not occur in other SM fields {\it that live on the D-brane}
because the centrifugal potential itself does not lead to a violation of Lorentz invariance.
For an open string stretching between revolving D-branes, 
the Coriolis force mixes radial and angular fluctuations, which
lead to the Lorentz violation in the Higgs dispersion relation.

In the next section,
 we consider a simple system of a D3-brane
 revolving around the origin of an six-dimensional extra space
 which is perpendicular to our three-dimensional space of the D3-brane world-volume.
The attractive potential for the D3-brane moduli fields 
 is simply assumed in the present paper. 
In section \ref{sec:violation},
 violation of the Lorentz symmetry in the Higgs sector 
 is discussed and we show that the dispersion relation of the Higgs field is modified. 
In section \ref{sec:higgs-mechanism},
we introduce the gauge symmetry and discuss 
 the Higgs mechanism in the presence of the Lorentz violation in the Higgs sector.
The last section is devoted to conclusion and discussions. 
 
\section{Higgs field on a revolving D-brane}
\label{sec:revolving}
The Higgs field is the only scalar field and the dynamics controls various important properties of the SM,
 especially  the electroweak symmetry breaking  and the generation of the fermion masses. 
 Nevertheless, we do not know much about the Higgs sector:
the origin of the Higgs potential, the hierarchy problem of the electroweak scale against various
UV scales, or the stability of the vacuum. 

In the following, we suppose that we live on a D3-brane in a ten-dimensional flat 
space-time\footnote{
It is straightforward to generalize the analysis to a curved background space,
 but the geometry between D-branes in very short distances, less than the string scale,
 will be encoded in the moduli space of vacua in D-brane world-volume theory
 \cite{Douglas:1996yp}. 
}. 
Standard model (SM) particles including fermions and gauge bosons are assumed to propagate on the brane.
Then what is the stringy interpretation of the Higgs field? 
In string theory, 
a scalar field may appear as a moduli field and its dynamics may be
described by geometrical properties of the  configuration. 
Here we consider a simple scenario that
 the Higgs field is a moduli field corresponding to the distance of D-branes,
 one at some point in the extra dimension, which we call the center,
 and the other revolving around the center on the $(X^8, X^9)$ plane
in the extra dimensional space. 

The embedding of a D3-brane in a ten-dimensional space-time
 is described by  $X^\mu(\xi)$ with $\mu=0,1,\cdots,9$. 
 They are the coordinates of the D-brane in the ten-dimensional spacetime
 and $\xi^a$ ($a=0,1,2,3$) are the coordinates of the D3-brane world-volume. 
We  take a gauge such that $X^a = \xi^a$ ($a=0, 1, 2, 3$) are satisfied. 
The revolution of the D3-brane around the origin 
is described as
\begin{eqnarray}
 X^8 &=& \phi^8 \cos \omega_0 \xi^0 - \phi^9 \sin \omega_0 \xi^0,
 \nonumber \\
 X^9 &=& \phi^9 \cos \omega_0 \xi^0 + \phi^8 \sin \omega_0 \xi^0, 
\label{moduli-rotating-coordinates}
\end{eqnarray}
 where $\omega_0$ is the angular frequency of the revolution of the D3-brane, 
 and determined later as a classical solution
 in the presence of appropriate attractive 
 force (or potential) between D (or anti-D)-branes.
 The fields $\phi^{8,9}(\xi)$ are the moduli fields on the D3-brane
 and represent the fluctuations of the D-brane in the rotating coordinate system. 
The low-energy effective action of the D3-brane moduli field is given by 
\begin{equation}
 S = - T_3 \int d^4 \xi \sqrt{-\det G_{ab}},
\end{equation}
 where $T_3$ is the D3-brane tension and $G_{ab}$ is the induced metric
\begin{equation}
 G_{ab} = \frac{\partial X^\mu}{\partial \xi^a} \frac{\partial X^\nu}{\partial \xi^b} \eta_{\mu\nu}.
\end{equation}
Ten-dimensional Minkowski metric is given by
 $\eta_{\mu\nu} = {\rm diag} (-1,1,\cdots,1)$
 and we neglect the B-field in the action for simplicity.
 Since we are interested in the dynamics of $X^8$ and $X^9$ fields, we set all the other irrelevant
 fields zero: $X^I=0$ ($I=4,5,6,7)$.
Then the induced metric of the embedding (\ref{moduli-rotating-coordinates}) is
given by 
\begin{equation}
 G_{ab} = \eta_{ab}
  + \left(\partial_a \phi^8 - \delta^0_a \omega_0 \phi^9 \right)
    \left(\partial_b \phi^8 - \delta^0_b \omega_0 \phi^9 \right)
  + \left(\partial_a \phi^9 + \delta^0_a \omega_0 \phi^8 \right)
    \left(\partial_b \phi^9 + \delta^0_b \omega_0 \phi^8 \right),
\end{equation}
 where  $\partial_a \equiv \partial/\partial\xi^a$.
 
If we neglect the fluctuations of the D-brane moduli field and 
set, e.g., $\phi^8=d$
  ($d$ is the distance of the D3-brane from the origin), 
 the effective action becomes
\begin{equation}
 S= -T_3 \int d^4 \xi \sqrt{1-(\omega_0 d)^2} 
 \sim - T_3 \int d^4 \xi (1 - \omega_0^2 d^2/2) .
\end{equation}
The second term is nothing but the centrifugal repulsive potential of the revolving D-brane.
Including the fluctuations and
 assuming 
$ \partial_a \phi^{8,9} \ll 1$ and 
$ \omega_0 \phi^{8,9} \ll 1$,
we can expand the action $S$ around $G_{ab} = \eta_{ab}$. 
At the lowest order (second order), the Lagrangian in the rotational coordinate system
becomes
\begin{equation}
 {\cal L} \simeq -T_3
 - \frac{1}{2} T_3
  \left( \partial^a \phi^8 \partial_a \phi^8 + \partial^a \phi^9 \partial_a \phi^9 \right)
 + \frac{1}{2} T_3 \omega_0^2 \left( (\phi^8)^2 + (\phi^9)^2 \right)
 - T_3 \omega_0 \left( \partial_0 \phi^8 \phi^9 - \phi^8 \partial_0 \phi^9 \right),
\end{equation}
 where $\partial^a \phi \partial_a \phi \equiv \eta^{ab} \partial_a \phi \partial_b \phi$.
The third term gives  the centrifugal potential and 
 and the forth term represents the Coriolis force, which mixies
radial and angular motions of the D-brane. 
In terms of the normalized polar-coordinate fields  $(R, \tilde{\Theta})$ defined by
\begin{equation}
 \sqrt{T_3} \left( \phi^8 + i \phi^9 \right)  = R e^{i \tilde{\Theta}},
\end{equation}
the Lagrangian is written as
\begin{eqnarray}
{\cal L} = -T_3 - \frac{1}{2} 
  \left\{ \partial^a R \partial_a R 
   + R^2 (\partial^a \tilde{\Theta} \partial_a \tilde{\Theta}) 
  -R^2 \omega_0^2 - 2 R^2 \omega_0 \partial_0 \tilde{\Theta}
   \right\}.
   \label{Lag-rotation}
\end{eqnarray}
The stationary motion of D-brane revolution is realized if an appropriate attractive potential $V(R)$
is generated for the radial direction $R$ of the moduli field. 
A simple example is
\begin{equation}
 V(R) = \frac{1}{2} \mu^2 R^2 + \frac{\lambda}{4} R^4.
 \label{V-R}
\end{equation} 
In a non-supersymmetric D-brane configuration, e.g., an anti-D-brane at the origin,
 the D3-brane is strongly attracted to the origin and we will obtain a potential such as (\ref{V-R}) 
 with a large coefficient $\mu \sim m_{\rm string}$ 
 \cite{Antoniadis:2000tq,Kitazawa:2012hr}.
Another different possibility for the potential will appear when we consider a 
 supersymmetric configuration. 
When D3-branes are at rest, there is no potential between D-branes since it is BPS
 and some supersymmetry remains. 
But if the D3-branes are in motion, 
 the supersymmetry is slightly broken, and very weak attractive force is generated
\cite{Bachas:1995kx,Lifschytz:1996iq,Douglas:1996yp,Iso-Ohta-Suyama}.
In such a case, the potential $V(R)$ depends not only on $R$ but also on $\omega_0$. 
In the present paper, 
an appropriate attractive potential is assumed that can 
balance the centrifugal repulsive force between revolving D-branes;
 we consider more general functions $V(R).$

In the effective Lagrangian (\ref{Lag-rotation}), we especially note that the Coriolis term is generated
which violates the Lorentz symmetry in the rotational frame. 
Hence, if Higgs is a geometrical moduli field and its vacuum expectation value
 is stabilized by revolution of D3-branes,
the Coriolis force induces  Lorentz-violating effects in the Higgs sector. It will be discussed
in the next section.
Also note that, in a rotationally invariant system,
  the angular momentum of the D3-brane is conserved
and the distance $R=R_0$ and 
the angular frequency  $\omega_0$ are correlated 
as $R_0^2 \omega_0 =I$ where $I$ is the angular momentum for a unit volume. 
Then the angular frequency is determined by the condition that $R_0 \omega_0^2 =V'(R_0)$.
For example,  if $V(R)$ is given by the specific function in (\ref{V-R}), the angular frequency is given by
$\omega_0^2=\mu^2 + \lambda R_0^2$. 


The  Lagrangian in the rotational frame (\ref{Lag-rotation})  can be, of course,  directly
derived from the ordinary Lagrangian in the inertial frame
\begin{equation}
 {\cal L} = 
  - \frac{1}{2}
    \left\{ \partial^a R \partial_a R + R^2 (\partial^a \Theta \partial_a \Theta) \right\} -V(R),
\label{ineretial-L}
\end{equation}
by replacing  $\Theta=\omega_0 \xi^0 +\tilde{\Theta}$.
Thus  $R = R_0$ and $\Theta = \omega_0 \xi^0$ are solutions to
the ordinary equations of motion
\begin{eqnarray}
 && \Box R - R \partial^a \Theta \partial_a \Theta - \partial_R V(R) =0, 
\nonumber \\
&&  \partial_a \left( R^2 \partial^a \Theta \right ) =0 .
\label{EOM-inertial}
\end{eqnarray}
In the following, we study the linearized equations of motion for the fluctuations,
$R=R_0+\tilde{R}$ and $\Theta=\omega_0+ \tilde{\Theta}$, around 
the classical solution.

\section{Lorentz Violation in the Higgs dispersion relations}
\label{sec:violation}
In the rotational frame, 
the linearized field equations around the classical solution are given by
\begin{equation}
 \ddot{\tilde{R}}
  - \nabla^2 \tilde{R} +  \omega_{\rm eff}^2 \tilde{R}
  - 2 \omega_0 R_0 \dot{\tilde{\Theta}} = 0,
\label{eq-radial}
\end{equation}
\begin{equation}
 R_0 \left( \ddot{\tilde{\Theta}} - \nabla^2 \tilde{\Theta} \right)
 + 2 \omega_0 \dot{\tilde{R}} = 0,
\label{eq-angular}
\end{equation}
 where 
 the dot means a derivative by $\xi^0$ and
 $\omega_{\rm eff}^2 = V''(R_0) -\omega_0^2$.
 For the potential (\ref{V-R}), $\omega_{\rm eff}^2=2 \lambda R_0^2$. 
 In general potential, these two frequencies $\omega_0$ and $\omega_{\rm eff}$ can be 
 chosen independently. 
Lorentz violating effects come from the Coriolis force;
the radial fluctuation $\tilde{R}$ and angular fluctuation $\tilde{\Theta}$ are mixed.
Thus in order to obtain the dispersion relation,  we need to diagonalize the equations of motion.
Introducing the Fourier integral representations
\begin{equation}
 \tilde{R}
  = \int d\omega d^3p \, R_{\omega,{\bf p}} \cos (\omega \xi^0 - {\bf p} \cdot {\bf \xi}),
\quad
 \tilde{\Theta}
  = \int d\omega d^3p \, \Theta_{\omega,{\bf p}} \sin (\omega \xi^0 - {\bf p} \cdot {\bf \xi}), 
\end{equation}
the field equations are written as 
\begin{equation}
\hat{L}  \left(
  \begin{array}{c}
   R_{\omega,{\bf p}} \\
   R_0 \Theta_{\omega,{\bf p}}
  \end{array}
 \right) 
=
 \left(
  \begin{array}{cc}
   \omega^2 - p^2 - \omega_{\rm eff}^2
   & 2 \omega_0 \omega  \\
   2 \omega_0 \omega & \omega^2 - p^2
  \end{array}
 \right)
 \left(
  \begin{array}{c}
   R_{\omega,{\bf p}} \\
   R_0 \Theta_{\omega,{\bf p}}
  \end{array}
 \right) = 0,
\end{equation}
 where $p \equiv |{\bf p}|$.
The dispersion relations can be obtained from the condition $\det \hat{L}=0$;
\begin{equation}
 \omega^2 - p^2 - \frac{\omega_{\rm eff}^2}{2}
 \pm \sqrt{ \left( \frac{\omega_{\rm eff}^2}{2}\right)^2
 + 4 \omega_0^2 \omega^2} = 0, 
\end{equation}
which is solved as 
\begin{equation}
 \omega_\pm^2
  = p^2 + M^2
    \pm M^2 \sqrt{1 + \frac{4 \omega_0^2 p^2}{M^4}},
    \hspace{10mm}
    M^2 \equiv 2\omega_0^2 + \frac{\omega_{\rm eff}^2}{2}  .
\label{disparsion-relations}
\end{equation}
A similar dispersion relation is discussed  \cite{Achucarro:2012sm}
 in the context of fluctuations around time-dependent background
 in cosmology.
For the potential (\ref{V-R}), $M^2= 2 m^2 + 3 \lambda R_0^2$.
The corresponding eigenvectors to the frequencies $\omega_\pm$ are given by
\begin{equation}
 V_{\pm} \propto 
 \left(
  \begin{array}{c}
 \omega_{\rm eff}^2
    \pm \sqrt{ \omega_{\rm eff}^4 + (4\omega_0\omega)^2} \\
   4 \omega_0 \omega
  \end{array}
 \right),
\end{equation}
 respectively. The Coriolis force mixes the radial component (the upper component) and the angular
 component (the lower component). Accordingly the Lorentz symmetry is broken.
 
In the small momentum region $\omega_0^2 p^2 \ll M^4$, 
the dispersion relations can be expressed as
\begin{eqnarray}
 \omega_+^2 &\simeq&
  \left( 1 + \frac{2\omega_0^2}{M^2} \right) p^2 + 2 M^2,
  \label{omega+}
\\
 \omega_-^2 &\simeq&
  \left( 1 - \frac{2\omega_0^2}{M^2} \right) p^2.
\label{Dispersion-relations}
\end{eqnarray}
Thus as far as the condition  $\omega_0^2 \ll M^2$, namely  
 $\omega_0^2 \ll \omega_{\rm eff}^2$, are satisfied,
the violation of Lorentz violation is small. 
The  eigenvectors  are respectively given by 
\begin{equation}
 V_+ \sim
  \left(
   \begin{array}{c}
    1 \\ \epsilon
   \end{array}
  \right),
\quad
 V_- \sim
  \left(
   \begin{array}{c}
    -\epsilon \\ 1
   \end{array}
  \right),
\end{equation}
 where $\epsilon$ is a small number.  Hence the $\omega_+$ mode is almost the Higgs mode
fluctuating along the radial direction while the other $\omega_-$ mode is the 
 angular mode. Due to the Coriolis force, they are slightly mixed.

The dispersion relations (\ref{Dispersion-relations}) may
indicate that the speeds of light 
are modified by the revolutions.
But its correct interpretation is the amplitude-modulation effect. 
Consider a circular polarized light 
 propagating in the $z$ direction;
\begin{equation}
 A_x = A \cos(\omega t - k z),
\quad
 A_y = \pm A \sin(\omega t - k z),
\end{equation}
 where $A$ is the amplitude and $\omega = c k=p$.
In a rotating coordinate frame we can define
\begin{eqnarray}
 A^{\rm rot}_1 &\equiv& A_x \cos(\omega_0 t) - A_y \sin(\omega_0 t)
  = A \cos((\omega \pm \omega_0)t - k z),
\\
 A^{\rm rot}_2 &\equiv&  A_x \sin(\omega_0 t) + A_y \cos(\omega_0 t) 
  = \pm A \sin((\omega \pm \omega_0)t - k z).
\end{eqnarray}
Now we see that the effective frequencies are modified as
 $\omega_\pm  =(\omega \pm \omega_0)$ and satisfies the modified dispersion 
 relations $\omega_\pm^2 =p^2 + \omega_0^2 \pm 2\omega_0 |p|.$
 The modification is  simply because the amplitudes oscillate with time in the rotating coordinate system. 
 The apparent light velocity seems to exceed $c=1$, but of course, 
 it does not mean that the causality is violated. 
 Similarly in our case, the fields in the inertial coordinate are massive with mass 
 $V''(R_0)=\omega_{\rm eff}^2+\omega_0^2$ and obey the dispersion relation
  $\omega^2 = p^2+(\omega_{\rm eff}^2+\omega_0^2$).
  Suppose $\omega_{\rm eff}=0$ such that there are no interactions in the inertial frame. 
 Then the effective frequencies $\omega_\pm$ satisfy
 \begin{equation}
 \omega_\pm^2 = (\omega \pm \omega_0)^2 = p^2 + 
 2\omega_0^2 
 \pm 2\omega_0 \sqrt{ p^2  +\omega_0^2}
\end{equation}
and the 
dispersion relations (\ref{disparsion-relations}) at $\omega_{\rm eff}=0$ are reproduced.


\section{Higgs mechanism in Lorentz violation}
\label{sec:higgs-mechanism}

In the previous section, we concentrated on a single D-brane revolving around the center
and showed that the Higgs field is mixed with the angular component 
by the Coriolis force which leads to the Lorentz violation in the dispersion relation. 
We now generalize the analysis to include gauge symmetry. 
Let us consider a stack of two parallel D3-branes. Then
the scalar fields $X^I$ on the brane  become $2 \times 2$ matrices 
\begin{equation}
 X^I = (X^I)^\alpha \frac{\tau^\alpha}{2},
\end{equation}
 where $\tau^{1,2,3}$ are Pauli matrices and $\tau^0$ is the unit matrix. 
The low-energy effective Lagrangian on the D3-branes
 is  represented by the Yang-Mills theory in four-dimensional spacetime. 
 As in the previous section, 
 we are interested in the motion of  D3-branes on 8-9 space. 
 The relevant part of the bosonic Lagrangian (setting $T_3=1)$ is now given by
\begin{equation}
 {\cal L}
  = - {\rm tr} \sum_{i=8,9}  D^\mu X^i D_\mu X^i  
    + g^2 {\rm tr} \left( \left[ X^8, X^9 \right]^2   \right)
    - \frac{1}{4} {\rm tr} \left( F^{\mu \nu} F_{\mu\nu} \right),
    \label{Lag-gauge}
\end{equation}
 where $\mu, \nu$ are summed over $0,1,2,3$ and $ A_\mu = A^{(\alpha)}_\mu \tau^\alpha/2$ are gauge fields
 on the branes whose field strength is given by 
 $F_{\mu \nu} = \partial_\mu A_\nu - \partial_\nu A_\mu - i g \left[ A_\mu, A_\nu\right]$.
 The scalar fields $X^I$ are in the adjoint representation of the $SU(2)$ gauge group
 and the covariant derivative is defined by
\begin{equation}
 D_\mu X^{i} = \partial_\mu X^{i} - i g \left[ A_\mu, X^{i} \right],
\end{equation}
 where $g$ is the gauge coupling constant.
We study the gauge symmetry breaking in terms of the rotational coordinates 
 $\phi^{8,9}$  defined by (\ref{moduli-rotating-coordinates}). 
Then the Lagrangian (\ref{Lag-gauge}) becomes
\begin{eqnarray}
 {\cal L} &=&
  - {\rm tr}\sum_i D^\mu X^i D_\mu X^i 
  - 2 \omega_0 {\rm tr} \left(  \phi^9 D_0 \phi^8  - \phi^8 D_0 \phi^9 \right)
\nonumber\\
&&
  + \omega_0^2 {\rm tr} \left( (\phi^8)^2 + (\phi^9)^2 \right)
  + g^2 {\rm tr} \left( \left[ \phi^8, \phi^9 \right]^2 \right)
   - \frac{1}{4} {\rm tr} \left( F^{\mu \nu} F_{\mu\nu} \right).
\label{SYM-rotatin-coordinates}
\end{eqnarray}
In order to balance the repulsive centrifugal potential, 
we add an appropriate attractive potential $V(R)$, where $R^2=\sum_{i=8,9} (\phi^i)^2$, 
such as  eq.(\ref{V-R}). 
Then the Lagrangian in the rotational coordinate frame is given by
\begin{equation}
 {\cal L}_{\rm eff} =
  - {\rm tr}  \left( D^\mu \phi^i D_\mu \phi^i \right)
  - 2 \omega_0 {\rm tr} \left( \phi^9 D_0 \phi^8  - \phi^8  D_0 \phi^9 \right)
  - V_{\rm eff}(R)
   - \frac{1}{4} {\rm tr} \left( F^{\mu \nu} F_{\mu\nu} \right),
\label{effective-L}
\end{equation}
 where
\begin{eqnarray}
 V_{\rm eff} (R) &=&
  - \omega_0^2 {\rm tr} R^2 +V(R)  + g^2 {\rm tr} \left( \left[ \phi^8, \phi^9 \right]^2 \right) .
 \label{effective-V}
\end{eqnarray}
Thus a new effect in the gauge theory is the coupling between the temporal component
 $A_0$ in the covariant derivative and the scalar fields $\phi^i$.  
 
 Let us first look at the spectrum of various fields without taking  
 Lorentz violating effects into account.
 Suppose that the classical solution  is given by
\begin{equation}
 \phi^8= R_0 \frac{\tau^3}{2},  \ \ \phi^9=0.
 \label{classical-solution}
\end{equation}
Thus fluctuations in the $\phi^8$ direction are radial modes while those in the $\phi^9$ direction
are angular modes along the revolution of the D3-brane.
There are  8 scalar degrees of freedom of $\phi^{i,(\alpha)}$ ($i=8,9$ and $\alpha=0,1,2,3$).
$U(2)$ gauge symmetry is spontaneously broken down to $U(1) \times U(1)$
and the W-bosons $A_\mu^{(\alpha)}$ ($\alpha=1,2)$ become massive with mass $m_A^2= g^2R_0^2$. 
The gauge boson $A_\mu^{(0)}$ and $A_\mu^{(3)}$ remain massless, which correspond
to the remaining $U(1) \times U(1)$ gauge symmetry.
Since the gauge transformation $\delta \phi^8 = i [G^a \tau^a/2, R_0 \tau^3/2]$ 
of the classical solution generates
$ \delta \phi^8 = G^1 \tau^2/2 - G^2 \tau^1/2$, we can identify two components
$\tilde{\phi}^{8,(i)}$ ($i=1,2$) as the would-be NG bosons and eaten by the longitudinal 
components of the gauge fields. In the Unitary gauge, we can set $\tilde{\phi}^{8,(i)}=0$ ($i=1,2$).
Other 2 scalar components, $\tilde{\phi}^{9,(i)}$ ($i=1,2$) acquire masses $m_S^2=g^2 R_0^2$
through the potential $g^2 {\rm tr} ([\phi^8, \phi^9]^2)$, where $\tilde{\phi}^9$
is fluctuation around  $\phi^9=0$, so $\tilde{\phi}^9=\phi^9.$
We will see  in the following that the mass $m_S^2$ of these scalar fields 
is modified by the Lorentz-violating effect of revolution. 
Another scalar field $\tilde{\phi}^{9,(3)}$ remains massless since it is a NG boson
associated with the global symmetry of the classical solution (\ref{classical-solution}) satisfying 
$(\phi^{8,(3)})^2 + (\phi^{9,(3)})^2 = R_0^2$. 

Now we investigate the effects of the Lorentz violation.
The Higgs mode is given by the fluctuation  $\tilde{\phi}^{8,(3)}$ around the classical solution,
$\phi^{8,(3)}=R_0 + \tilde{\phi}^{8,(3)}$, 
and the Lorentz violation in the dispersion relation is calculated as in the previous section. 
The Coriolis term in the rotational frame Lagrangian, which is linear in $\omega_0$, 
is now given in the Unitary gauge $\tilde{\phi}^{8,(i)}=0$ ($i=1,2$) as
\begin{eqnarray}
{\cal L}_{\rm Coriolis} = \omega_0  \sum_{i=0,3} \left( 
 \tilde{\phi}^{8,(i)} \partial_t \tilde{\phi}^{9,(i)} -   \tilde{\phi}^{9,(i)} \partial_t \tilde{\phi}^{8,(i)} \right)
   - 2 g \omega_0 R_0 \left( A_0^{(1)} \tilde{\phi}^{9,(2)}  -   A_0^{(2)} \tilde{\phi}^{9,(1)} \right)
\end{eqnarray}
where we have dropped a total derivative term 
and terms cubic in fluctuations. 
The $i=3$ component of the first term is the mixing between the Higgs $\tilde{\phi}^{8,(3)}$
and the angular mode $\tilde{\phi}^{9,(3)}$ and already investigated in the previous section.
An interesting term is the mixing between the temporal component of the W-bosons $A_0^{(i)}$
and the massive scalar field $\tilde{\phi}^{9,(i)}$ ($i=1,2$) with mass $m_S=gR_0$.
Since the temporal component of gauge fields is not dynamical, we can integrate
over $A_0^{(i)}$. In the present case, since the gauge boson
is massive with mass $m_A$, the potential is Yukawa-type. 
The relevant part of the Lagrangian for $A_0^{(1)}$ is 
\begin{equation}
 \frac{1}{2} A_0^{(1)} (\Delta -m_A^2) A_0^{(1)} -g A_0^{(1)} \rho^{(1)}(x)
\end{equation}
where the charge density $\rho^{(1)}$ is modified by the effect of revolution;
\begin{equation}
 \rho^{(1)} = \nabla \cdot  \dot{{\bf A}}^{(1)} + 2 \omega_0 R_0 \tilde{\phi}^{9,(2)} .
\end{equation}
The self-interaction terms of gauge fields are neglected. 
The second term of $\rho^{(1)}$ can be interpreted as an induced charge density associated with 
the revolution of D3-brane. 
An integration over $A_0^{(1)}$ gives the non-local interaction $g^2 \rho^{(1)} (\Delta -m_A^2)^{-1} \rho^{(1)}$
and gives a $\omega_0$-dependent correction to the mass $m_S^2$;
\begin{equation}
 m_S^2 = g^2 R_0^2 + \frac{4g^2\omega_0^2 R_0^2}{m_A^2} = g^2 R_0^2 + 4 \omega_0^2 .
\end{equation}
But, for phenomenological reason, we are interested in the region $\omega_0^2 \ll g^2 R_0^2 =m_A^2$
and the correction is not so large.

\section{Conclusions and Discussions}
\label{sec:conclusion}

After the discovery of the Higgs boson, 
 our view towards the physics beyond the standard model needs reconsiderations.
We have not discovered any deviations from the standard model, but
many important issues in particle physics are left unsolved, including 
dark matter, baryon asymmetry of the universe and dark energy;
and the most urgent issue will be the origin of the electroweak symmetry breaking. 
The hierarchy problem, the stability of the
electroweak symmetry breaking scale against various UV energy scales,
will be a key to go beyond the standard model. 
There are many interesting proposals to solve the hierarchy problem, 
both in the bottom-up and in the top-down approaches,
but there is no definitive solution yet.  
In such a situation, an unconventional idea about the Higgs boson
 will be also worth being investigated.

In the present paper, we study a possibility of Lorentz violation in the Higgs sector. 
The idea behind is very simple. In string theory, we may live on a D3-brane embedded
in a higher dimensional space-time and Higgs field is 
a moduli field representing a distance between D-branes. 
Then the hierarchy problem is reinterpreted as a geometrical stability of the
D-brane configurations in higher dimensional space. 
Like our solar system, the configuration may be stabilized by a stationary motion of D-branes,
in particular revolution of D3-brane with an angular frequency $\omega_0$.
If it is the case, Lorentz symmetry will be violated. 
The purpose of the present paper is to study possible consequences of such
Lorentz violating effects. 
We show that the dispersion relation of the Higgs boson is modified through the Coriolis force. 
If the angular frequency is smaller than the Higgs mass, Lorentz violation is tiny and
the Higgs dispersion relation (\ref{omega+}) is given by 
\begin{equation}
 \omega^2 = \left(1+ \frac{4 \omega_0^2}{M_H^2} \right) p^2 + M_H^2 .
\end{equation}
From the Higgs boson experiments,
 we can get an upper bound of the angular frequency $\omega_0$. 
A produced Higgs boson decays into two neutral gauge bosons, $Z$ and $Z^*$, and then into 4 leptons. 
The recent data from ATLAS experiment \cite{ATLAS} shows that the mass of the Higgs 
boson is determined as $124.79 \pm 0.37$ GeV from the 4 lepton decay channel. 
Since the Higgs boson momenta $p$
vary event by event, the Lorentz-violating effect $4\omega_0^2 p^2/M_H^2$ 
also varies and it is misinterpreted as
 the statistical error for the Higgs boson mass. 
Suppose that the momentum variation is of the similar order of the Higgs boson mass \cite{ATLAS2}, 
the small experimental error $0.37$ GeV suggests that the angular frequency $\omega_0$
must follow 
\begin{equation}
 \omega_0 \lesssim {\cal O}(0.1) {\rm GeV}.
\end{equation}
Thus we already have a stringent constraint on $\omega_0$. 

In the present paper, we simply assume that an appropriate attractive potential
is generated to balance the centrifugal potential. 
If we start from a non-BPS configuration, a very strong attractive potential 
is generated and the typical angular frequency $\omega_0$ is given by
the string scale \cite{Iso:2015mva}, which is already excluded in the Higgs experiments. 
Furthermore such a stationary motion
is unstable by emitting gravitons or other closed string particles. 
Thus in order to utilize the mechanism of revolution to stabilize the moduli field,
 it will be necessary to start from a BPS configuration, such as two parallel 
Dp-branes. When they are at rest, there is no potential and the static configuration
is stable: the moduli potential is flat.  
But when they move each other, supersymmetry is broken and an attractive
interaction is induced \cite{Bachas:1995kx,Lifschytz:1996iq,Douglas:1996yp}. 
It is now tempting to ask whether a bound state (or a resonant state) 
can be formed whose distance is much shorter than the string scale. 
In the D0-brane case, several previous studies 
\cite{Kabat:1996cu, Danielsson:1996uw} suggest an existence of a resonant state. 
Higher-dimensional cases are not much studied. 
Besides an application to the electroweak symmetry breaking, 
D-branes in motion will play an important role 
in the string cosmology, especially a stringy realization of the primordial inflation
 \cite{Dvali:1998pa,Kehagias:1999vr, Silverstein:2003hf}. 
It addition,  importance of angular motion of D-branes  is pointed out in \cite{Easson:2007dh}.
Nevertheless D-brane dynamics under motion are not yet fully understood. 
We want to come back to these problems in future investigations. 

\section*{Acknowledgments}

The authors would like to thank Takao Suyama and Hikaru Ohta for discussions.
This work is supported in part by Grant-in-Aid for Scientific Research
(18H03708, 16H06490 and 19K03851)  from MEXT Japan.
NK would like to thank KEK theory center for the kind hospitality.


\begin{thebibliography}{99}


\bibitem{Blumenhagen:2006ci}
  R.~Blumenhagen, B.~Kors, D.~Lust and S.~Stieberger,
  Phys.\ Rept.\  {\bf 445} (2007) 1
  [hep-th/0610327].
\bibitem{Ibanez:2012zz}
  L.~E.~Ibanez and A.~M.~Uranga,
  ``String theory and particle physics: An introduction to string phenomenology,''
  Cambridge University Press, Cambridge U.K. (2012).

\bibitem{Dvali:1998pa}
  G.~R.~Dvali and S.~H.~H.~Tye,
  Phys.\ Lett.\ B {\bf 450} (1999) 72
  [hep-ph/9812483].
\bibitem{Kehagias:1999vr}
  A.~Kehagias and E.~Kiritsis,
  JHEP {\bf 9911} (1999) 022
  [hep-th/9910174].
\bibitem{Silverstein:2003hf}
  E.~Silverstein and D.~Tong,
  Phys.\ Rev.\ D {\bf 70} (2004) 103505
  [hep-th/0310221].
\bibitem{Easson:2007dh}
  D.~A.~Easson, R.~Gregory, D.~F.~Mota, G.~Tasinato and I.~Zavala,
  JCAP {\bf 0802} (2008) 010
  [arXiv:0709.2666 [hep-th]].

\bibitem{Iso:2015mva}
  S.~Iso and N.~Kitazawa,
  PTEP {\bf 2015} (2015) no.12,  123B01
  [arXiv:1507.04834 [hep-ph]].
 
\bibitem{Antoniadis:2000tq}
  I.~Antoniadis, K.~Benakli and M.~Quiros,
  Nucl.\ Phys.\ B {\bf 583} (2000) 35
  [hep-ph/0004091].
\bibitem{Kitazawa:2012hr}
  N.~Kitazawa and S.~Kobayashi,
  Phys.\ Lett.\ B {\bf 720} (2013) 373
  [arXiv:1211.1777 [hep-th]].
  
\bibitem{Kabat:1996cu}
  D.~N.~Kabat and P.~Pouliot,
  Phys.\ Rev.\ Lett.\  {\bf 77}, 1004 (1996)
  [hep-th/9603127].
\bibitem{Danielsson:1996uw} 
  U.~H.~Danielsson, G.~Ferretti and B.~Sundborg,
  Int.\ J.\ Mod.\ Phys.\ A {\bf 11}, 5463 (1996)
  [hep-th/9603081].

\bibitem{Iso-Ohta-Suyama}
  S.~Iso, H.~Ohta and T.~Suyama, 
  arXiv:1812.11505 [hep-th].

\bibitem{LV}
 D.~Mattingly,
  Living Rev.\ Rel.\  {\bf 8}, 5 (2005)
  [gr-qc/0502097].
 R.~Bluhm,
  Lect.\ Notes Phys.\  {\bf 702}, 191 (2006)
  [hep-ph/0506054]; 
 V.~A.~Kostelecky and N.~Russell,
  Rev.\ Mod.\ Phys.\  {\bf 83}, 11 (2011)
  [arXiv:0801.0287 [hep-ph]].

  
\bibitem{LV-Higgs}
 D.~L.~Anderson, M.~Sher and I.~Turan,
  Phys.\ Rev.\ D {\bf 70}, 016001 (2004)
  [hep-ph/0403116].
 R.~Lehnert,
  J.\ Phys.\ Conf.\ Ser.\  {\bf 952}, no. 1, 012008 (2018).
 S.~Aghababaei, M.~Haghighat and A.~Kheirandish,
  Phys.\ Rev.\ D {\bf 87}, no. 4, 047703 (2013)
  [arXiv:1302.5023 [hep-ph]].
  B.~Altschul,
  Phys.\ Rev.\ D {\bf 86}, 045008 (2012)
  [arXiv:1202.5993 [hep-th]].
 
\bibitem{Douglas:1996yp}
  M.~R.~Douglas, D.~N.~Kabat, P.~Pouliot and S.~H.~Shenker,
  Nucl.\ Phys.\ B {\bf 485} (1997) 85
  [hep-th/9608024].
\bibitem{Bachas:1995kx}
  C.~Bachas,
  Phys.\ Lett.\ B {\bf 374} (1996) 37
  [hep-th/9511043].
\bibitem{Lifschytz:1996iq}
  G.~Lifschytz,
  Phys.\ Lett.\ B {\bf 388} (1996) 720
  [hep-th/9604156].


\bibitem{Achucarro:2012sm}
  A.~Achucarro, J.~O.~Gong, S.~Hardeman, G.~A.~Palma and S.~P.~Patil,
  JHEP {\bf 1205} (2012) 066
  [arXiv:1201.6342 [hep-th]].



\bibitem{ATLAS}
  M.~Aaboud {\it et al.} [ATLAS Collaboration],
  Phys.\ Lett.\ B {\bf 784} (2018) 345
  [arXiv:1806.00242 [hep-ex]].

\bibitem{ATLAS2}
  M.~Aaboud {\it et al.} [ATLAS Collaboration],
  JHEP {\bf 1803} (2018) 095
  [arXiv:1712.02304 [hep-ex]].

\end{thebibliography}
\end{document}